\documentstyle[12pt]{article}

\catcode`\@=11
\long\def\@makefntext#1{
\protect\noindent \hbox to 3.2pt {\hskip-.9pt
$^{{\ninerm\@thefnmark}}$\hfil}#1\hfill}		%CAN BE USED

 \def\@makefnmark{\hbox to 0pt{$^{\@thefnmark}$\hss}}  %ORIGINAL

\def\ps@myheadings{\let\@mkboth\@gobbletwo
\def\@oddhead{\hbox{}
\rightmark\hfil\ninerm\thepage}
\def\@oddfoot{}\def\@evenhead{\ninerm\thepage\hfil
\leftmark\hbox{}}\def\@evenfoot{}
\def\sectionmark##1{}\def\subsectionmark##1{}}

\newcounter{sectionc}\newcounter{subsectionc}\newcounter{subsubsectionc}
\renewcommand{\section}[1] {\vspace{0.6cm}\addtocounter{sectionc}{1}
\setcounter{subsectionc}{0}\setcounter{subsubsectionc}{0}\noindent
	{\bf\thesectionc. #1}\par\vspace{0.4cm}}
\renewcommand{\subsection}[1] {\vspace{0.6cm}\addtocounter{subsectionc}{1}
	\setcounter{subsubsectionc}{0}\noindent
	{\it\thesectionc.\thesubsectionc. #1}\par\vspace{0.4cm}}
\renewcommand{\subsubsection}[1]
{\vspace{0.6cm}\addtocounter{subsubsectionc}{1}
	\noindent {\rm\thesectionc.\thesubsectionc.\thesubsubsectionc.
	#1}\par\vspace{0.4cm}}

\newcounter{appendixc}
\newcounter{subappendixc}[appendixc]
\newcounter{subsubappendixc}[subappendixc]

\renewcommand{\appendix}[1] {\vspace{0.6cm}
        \refstepcounter{appendixc}
        \setcounter{figure}{0}
        \setcounter{table}{0}
        \setcounter{equation}{0}
        \renewcommand{\thefigure}{\Alph{appendixc}.\arabic{figure}}
        \renewcommand{\thetable}{\Alph{appendixc}.\arabic{table}}
        \renewcommand{\theappendixc}{\Alph{appendixc}}
        \renewcommand{\theequation}{\Alph{appendixc}.\arabic{equation}}
        \noindent{\bf Appendix \theappendixc #1}\par\vspace{0.4cm}}

\def\abstracts#1{{
	\centering{\begin{minipage}{30pc}\tenrm\baselineskip=12pt\noindent
	\centerline{\tenrm ABSTRACT}\vspace{0.3cm}
	\parindent=0pt #1
	\end{minipage}}\par}}

\renewenvironment{thebibliography}[1]
	{\begin{list}{\arabic{enumi}.}
	{\usecounter{enumi}\setlength{\parsep}{0pt}
\setlength{\leftmargin 1.25cm}{\rightmargin 0pt}
	 \setlength{\itemsep}{0pt} \settowidth
	{\labelwidth}{#1.}\sloppy}}{\end{list}}

\newcounter{itemlistc}
\newcounter{romanlistc}
\newcounter{alphlistc}
\newcounter{arabiclistc}

\newcommand{\fcaption}[1]{
        \refstepcounter{figure}
        \setbox\@tempboxa = \hbox{\tenrm Fig.~\thefigure. #1}
        \ifdim \wd\@tempboxa > 6in
           {\begin{center}
        \parbox{6in}{\tenrm\baselineskip=12pt Fig.~\thefigure. #1}
            \end{center}}
        \else
             {\begin{center}
             {\tenrm Fig.~\thefigure. #1}
              \end{center}}
        \fi}

\newcommand{\tcaption}[1]{
        \refstepcounter{table}
        \setbox\@tempboxa = \hbox{\tenrm Table~\thetable. #1}
        \ifdim \wd\@tempboxa > 6in
           {\begin{center}
        \parbox{6in}{\tenrm\baselineskip=12pt Table~\thetable. #1}
            \end{center}}
        \else
             {\begin{center}
             {\tenrm Table~\thetable. #1}
              \end{center}}
        \fi}

\def\@citex[#1]#2{\if@filesw\immediate\write\@auxout
	{\string\citation{#2}}\fi
\def\@citea{}\@cite{\@for\@citeb:=#2\do
	{\@citea\def\@citea{,}\@ifundefined
	{b@\@citeb}{{\bf ?}\@warning
	{Citation `\@citeb' on page \thepage \space undefined}}
	{\csname b@\@citeb\endcsname}}}{#1}}

\newif\if@cghi
\def\cite{\@cghitrue\@ifnextchar [{\@tempswatrue
	\@citex}{\@tempswafalse\@citex[]}}
\def\citelow{\@cghifalse\@ifnextchar [{\@tempswatrue
	\@citex}{\@tempswafalse\@citex[]}}
\def\@cite#1#2{{$\null^{#1}$\if@tempswa\typeout
	{IJCGA warning: optional citation argument
	ignored: `#2'} \fi}}

\def\fnt#1#2{\footnotetext{\kern-.3em
	{$^{\mbox{\sevenrm #1}}$}{#2}}}

 1
 1
 1

\font\tenbf=cmbx10
\font\tenrm=cmr10
\font\tenit=cmti10

\font\ninerm=cmr9

\begin{document}

%\centerline{\tenbf QUARK CONFINEMENT AND THE HADRON SPECTRUM}
\baselineskip=22pt
\centerline{\tenbf COLOR CONFINEMENT AND DYNAMICAL EFFECT OF
LIGHT QUARKS}
\baselineskip=16pt
\centerline{\tenbf IN THE DUAL GINZBURG-LANDAU THEORY}
%\centerline{\ninerm (For 20\% Reduction to 8.5 $\times$ 6 in Trim Size)}
\vspace{0.8cm}
\centerline{\tenrm HIDEO~SUGANUMA}
\baselineskip=13pt
\centerline{\tenit RIKEN, Wako, Saitama 351-01, Japan}
%\baselineskip=12pt
%\centerline{\tenit City, State ZIP/Zone, Country}
\vspace{0.3cm}
\centerline{\tenrm and}
\vspace{0.3cm}
\centerline{\tenrm SHOICHI~SASAKI and HIROSHI~TOKI}
\baselineskip=13pt
\centerline{\tenit RCNP, Osaka University, Ibaraki, Osaka 567, Japan}
\vspace{0.9cm}
\abstracts{
We study nonperturbative features of QCD
using the dual Ginzburg-Landau theory.
The color confinement is realized through the dual Higgs mechanism,
which is brought by QCD-monopole condensation.
We investigate the infrared screening effect
on the color confinement due to the light-quark pair creation.
By solving the Schwinger-Dyson equation,
we find that the dynamical chiral-symmetry breaking
is largely brought by the confining force.
}

\vfil
\rm\baselineskip=14pt
\section{Dual Higgs Mechanism for Color Confinement}
{We study color confinement and dynamical effects of light quarks
in the dual Ginzburg-Landau (DGL) theory.\cite{{1.},{2.}}
Because of the asymptotic freedom,
QCD in the infrared region exhibits the nonperturbative features
like the color confinement and the dynamical chiral-symmetry
breaking (D$\chi $SB).
The color confinement is characterized by the vanishing
of the color dielectric constant and squeezing of the
color electric flux, so that
it has been regarded as the dual version of the
superconductor using the duality of gauge theories.
In this analogy, the confinement is brought by the dual
Meissner effect originated from QCD-monopole condensation,
which corresponds to Cooper-pair condensation in the superconductivity.
As for the appearance of monopoles in QCD,
't~Hooft\cite{3.} proposed an interesting idea of the abelian gauge
fixing, which is defined by the diagonalization of a
suitable gauge dependent variable. In this gauge, QCD is reduced into
an abelian gauge theory with magnetic monopoles,
which appear from the hedgehog-like configuration corresponding to
the nontrivial homotopy class on the nonabelian manifold,
$\pi _2({\rm SU}(N_c)/{\rm U}(1)^{N_c-1})=Z_\infty ^{N_c-1}$.

We compare the QCD vacuum with the superconductor
in terms of the abelian gauge fixing.
In the superconductor, there are two kinds of degrees of freedom,
the gauge field (photon) and the matter field corresponding to
the electron and the metallic lattice, which provide the Cooper pair.
On the other hand, there is only the gauge field in the pure gauge QCD,
and therefore it seems difficult to find the analogous point between
these two systems.
However, in the abelian gauge,
the diagonal part and the off-diagonal
part of gluons play different roles.
While the diagonal part behaves as the gauge field,
the off-diagonal part behaves as the charged matter and provides
QCD-monopoles, whose condensation leads to
the dual Higgs mechanism, mass generation of the dual gauge field.
Thus, QCD can be regarded as a similar system to the
dual superconductor in the abelian gauge.
Recent studies\cite{4.} using the lattice gauge theory
have reported many evidences on
the abelian dominance scheme and monopole condensation for the color
confinement.}

\section{Dual Ginzburg-Landau Theory and Quark Confinement Potential}
{The DGL theory is considered as an infrared
effective theory of QCD in the abelian gauge.
Its Lagrangian in the Zwanziger form is described
by the diagonal gluon $\vec A_\mu $ and
the dual gauge field $\vec B_\mu $,\cite{{1.},{2.}}
\begin{eqnarray}
{\cal L}_{\rm DGL}
&=&-{1 \over n^2}
[n\cdot (\partial \wedge \vec A)]^\nu
[n\cdot ^*(\partial \wedge \vec B)]_\nu
-{1 \over 2n^2}
[n\cdot (\partial \wedge \vec A)]^2
-{1 \over 2n^2}
[n\cdot (\partial \wedge \vec B)]^2 \cr
&&+\sum_{\alpha =1}^3 \left[|(i\partial_\mu -g \vec \epsilon _\alpha  \cdot
\vec B_\mu )\chi _\alpha |^2
-\lambda (|\chi _\alpha |^2-v^2)^2 \right]
\end{eqnarray}
apart from the quark sector.
(The notations are the same as those in Refs.1 and 2.)
The self-interaction of the QCD-monopole field
$\chi _\alpha $ is introduced phenomenologically like
the Ginzburg-Landau theory in the superconductivity.
There is the dual gauge symmetry corresponding to the
local phase invariance of the QCD-monopole field $\chi _\alpha $
as well as the residual gauge symmetry embedded in SU(3)$_c$.
When QCD-monopoles are condensed, the dual Higgs mechanism occurs
accompanying mass generation of the dual gauge field $\vec B_\mu$
and the spontaneous breaking of the dual gauge symmetry.
On the other hand, the residual gauge symmetry is never broken in this process.

In this framework, the Dirac condition $eg=4\pi $ for the
dual gauge coupling constant $g$ is naturally
derived in the same way as in the Grand Unified Theory.\cite{{1.},{3.}}
In view of the renormalization group,
the DGL theory is not
asymptotically free in terms of $g$ similar to the scalar QED.
Hence, asymptotic freedom is expected for the
gauge coupling constant $e$
owing to the Dirac condition.
Thus, the DGL theory qualitatively
shows asymptotic freedom in terms of $e$,\cite{1.}
which seems a desirable feature for an effective theory of QCD.

First, we investigate the $Q$-$\bar Q$ system
in the quenched level using the DGL theory.
The $Q$-$\bar Q$ static potential includes the Yukawa and linear
part,\cite{1.}
\begin{equation}
V(r)=-{\vec Q^2 \over 4\pi }\cdot {e^{-m_Br} \over r}+kr,
\quad
k={\vec Q^2 m_B^2 \over 8\pi }\ln({m_B^2+m_\chi ^2 \over m_B^2}),
\end{equation}
where $\vec Q$ denotes the color electric charge of the color source.
Here, $m_B$ is the mass of the dual gauge field $\vec B_\mu $, whose inverse
corresponds to the cylindrical radius of the flux tube.
The expression of the string tension $k$ is quite similar
to the energy per unit length of the Abrikosov
vortex in the type-II superconductor.\cite{1.}
}

\section{Quark Pair Creation, Infrared Screening Effect
and D$\chi $SB}

{Next, we consider the dynamical effect of light quarks,
which should be taken into account for the study of D$\chi $SB.
A long hadron string can be cut through the light $q$-$\bar q$
pair creation,
which is estimated by using the Schwinger formula.\cite{1.}
This provides the screening effect on the long-range part of
the confinement potential, as is observed in the lattice QCD
with light dynamical quarks.\cite{5.}
Taking account of such an infrared screening effect,
we introduce the corresponding infrared cutoff $\varepsilon $ to the
gluon propagator as\cite{1.}
\begin{equation}
D_{\mu \nu}^{sc}=-{1 \over k^2}
\left\{ g_{\mu \nu }+(\alpha_e-1)
{k_\mu k_\nu \over k^2} \right\}
-{1 \over k^2}
{m_B^2 \over k^2-m_B^2} \cdot
{\epsilon ^\lambda  \ _{\mu \alpha \beta }\epsilon _{\lambda \nu \gamma \delta
}n^\alpha n^\gamma k^\beta k^\delta
\over (n \cdot k)^2+\varepsilon ^2}
\end{equation}
without breaking the residual gauge symmetry.
Here, we have introduced the infrared cutoff
to the non-local factor ${1 \over (n \cdot k)^2}$,
which provides the strong and long-range correlation as
the origin of the confinement potential.
Using this gluon propagator,
we obtain a compact formula for the screened quark potential,\cite{1.}
\begin{equation}
V^{sc}_{\rm linear}(r)=k \cdot {1-e^{-\varepsilon r} \over \varepsilon }
\end{equation}
apart from the Yukawa part.
This screened potential certainly exhibits the saturation
for the longer distance than $\varepsilon ^{-1}$.

Finally, we investigate D$\chi$SB in the DGL theory.
We use the Schwinger Dyson (SD) equation with
the gluon propagator including the nonperturbative effects on the
confinement and the infrared screening.
We find that QCD-monopole condensation largely
contributes to the dynamical generation of the quark mass.\cite{1.}
As the physical interpretation,
the dual Higgs mechanism leads to the confining force
(a strong and long-range attractive force) between the
$q$-$\bar q$ pair with opposite color charges,
which promotes $q$-$\bar q$ pair condensation similarly in the
Nambu-Jona-Lasinio model.

In conclusion, we have studied nonperturbative features of QCD
using the dual Ginzburg-Landau theory.
The confinement potential has been reproduced
through the dual Higgs mechanism, which is brought by QCD-monopole
condensation.
We have investigated the infrared screening effect
on the linear confinement potential
due to the light-quark pair creation, and obtained
a compact formula for the screened potential.
D$\chi$SB have been also studied in terms of the SD equation.
We have found that D$\chi$SB is largely brought by the confining
force between the light $q$-$\bar q$ pair.

One of the authors (H.S.) is supported by the Special Researchers'
Basic Science Program at RIKEN.}


\vspace{0.4cm}
\noindent
{\bf REFERENCES}
\vspace{1pt}
\begin{thebibliography}{9}

\bibitem{1.}
%H.~Suganuma, S.~Sasaki and H.~Toki, preprint, RIKEN-AF-NP-164 (1994),
%to be published in {\it Nucl.~Phys.} {\bf B}.
H.~Suganuma, S.~Sasaki and H.~Toki,
to be published in {\it Nucl.~Phys.} {\bf B}.

\bibitem{2.}
T.~Suzuki, {\it Prog.~Theor.~Phys.} {\bf 80} (1988) 929 ;
{\bf 81} (1989) 752.

S.~Maedan and T.~Suzuki, {\it Prog.~Theor.~Phys.} {\bf 81} (1989) 229.
\bibitem{3.}
G.~'t~Hooft, {\it Nucl.~Phys.}~{\bf B190} (1981) 455.

\bibitem{4.}
A.~S.~Kronfeld, G.~Schierholz and U.-J.~Wiese, {\it Nucl.~Phys.}
{\bf B293} (1987) 461.

T.~Suzuki and I.~Yotsuyanagi, {\it Phys.~Rev.} {\bf D42} (1990) 4257.

\bibitem{5.}
W.~B\"urger, M.~Faber, H.~Markum, M.~M\"uller,
{\it Phys.~Rev.}~{\bf D47} (1993) 3034.

\end{thebibliography}
\end{document}